\documentclass[aps,prc,preprint,groupedaddress,showpacs]{revtex4-1}
\usepackage{graphicx}

\begin{document}

\title{Nuclear Structure Uncertainties in Parity-Violating Electron Scattering from Carbon 12}

\author{O. Moreno}
\author{T. W. Donnelly}
\affiliation{Center for Theoretical Physics, Laboratory for Nuclear Science and
  Department of Physics, Massachusetts Institute of Technology, Cambridge, MA 02139, USA}

\date{\today}

\begin{abstract}
High precision measurements of the parity-violating asymmetry in polarized electron scattering from nuclei can be used to extract information on nuclear and nucleon structure or to determine Standard Model couplings and higher-order radiative corrections. To this end, low uncertainties are also required in the effects that inevitably arise from modeling the underlying nuclear structure. An experimental precision of a few tenths of a percent may be attainable for the asymmetry if the appropriate kinematic range is chosen, as will be discussed here for the case of $^{12}$C. And given this, the dual goal of ascertaining both the sizes of various nuclear structure related effects and of providing estimates of their uncertainties for this particular target will be discussed.
\end{abstract}


\maketitle


\section{Introduction}\label{introduction}

Recently there has been interest expressed in having relatively low energy, high luminosity polarized electron beams for studies of parity-violating (PV) electron scattering \cite{MITworkshop}. One such is the MESA accelerator at Mainz \cite{aul11}; another might be an upgraded version of the FEL facility at Jefferson Lab \cite{nei00}. Various motivations underlie these initiatives including improved measurements of PV electron-proton scattering at low momentum transfers, of the neutron radii of nuclei and of tests of the Standard Model, both of the weak mixing angle and of higher order radiative corrections including single-nucleon box and cross-box two-boson exchange diagrams and dispersion corrections. Here PV scattering from nuclei is involved and this raises the questions: How well can nuclear structure effects be taken into account? What level of uncertainty exists in evaluating these nuclear effects, both at present or through future theoretical studies?

In the present paper we report the results of a study of specific classes of nuclear effects to be discussed in the following section, and indicate in some cases where additional work might be undertaken. The paper is organized in the following way: following these brief introductory comments, in Sec. \ref{formalism} the basic formalism is summarized. References are given in that section to much more detailed treatments of the formalism and so here only a few specifics needed for the present study are highlighted. The various effects that either stem from nuclear structure issues or involve strangeness content in the nucleons in the nucleus are enumerated at this point. Following this in Sec. \ref{results} results are given for the case of $^{12}$C, both for the assumed kinematics of the MESA facility and for somewhat higher energies that might become available with an upgraded JLab FEL facility. Included there is a careful analysis of the expected fractional uncertainties in the PV asymmetry expected with these, together with the uncertainties projected to arise from nuclear structure uncertainties and from our current knowledge of strangeness. The paper then concludes with a brief summary of our findings.


\section{Formalism}\label{formalism}

The weak interaction contains vector and axial-vector components having opposite behavior under a parity transformation. This is at the origin of the non-zero value of the PV asymmetry in electron scattering, which is defined as the relative difference between the cross sections of
incoming electrons longitudinally polarized parallel ($\sigma^+$) and antiparallel ($\sigma^-$) to their momentum:
\begin{equation}\label{asymmetry_sigmas}
A=\frac{d\sigma^+ - d\sigma^-}{d\sigma^+ + d\sigma^-}\, .
\end{equation}
By considering the exchange of a single gauge boson for each of the two interactions involved in the process, namely the neutral weak ($Z^0$ boson) and the electromagnetic (photon) interactions, and neglecting the effect of the nuclear Coulomb field on the electron wave functions, {\it i.e.} within plane wave Born approximation (PWBA), the PV asymmetry can be written as \cite{don89}
\begin{equation}\label{asymmetry_PWBA}
A=\frac{G_F \:|Q^2|}{2\pi\alpha\sqrt{2}}\:\frac{W^{PV}}{W^{PC}}\, .
\end{equation}
It is apparent that the PV asymmetry factorizes into a Standard-Model part, containing the Fermi (weak) and the fine-structure (electromagnetic) coupling constants, $G_F$ and $\alpha$ respectively, a four-momentum transfer dependence $|Q^2|$, and a nuclear-structure dependent part containing the ratio of the PV to the parity-conserving (PC) responses. The former arises from an electromagnetic - weak interaction interference (indicated with a hat in the hadronic tensors below) and contains terms with vector-vector tensors weighted by the weak neutral axial coupling of the electron $a_A^e$, both longitudinal ($L$) and transverse ($T$), as well as a term ($T'$) with axial-vector tensors weighted by the weak neutral vector coupling of the electron $a_V^e$:
\begin{equation}\label{WPV}
W^{PV} = a_A^e (v_L \hat{W}_L + v_T \hat{W}_T) + a_V^e v_{T'} \hat{W}_{T'} .
\end{equation}
The PC response, on the other hand, is purely electromagnetic and therefore contains just vector-vector tensors, both longitudinal and transverse:
\begin{equation}\label{WPC}
W^{PC} = v_L W_L + v_T W_T .
\end{equation}

When considering an $N=Z$ nuclear target with pure isospin $T=0$ in its ground state, only the isoscalar part of the elastic responses is involved, which effectively removes the axial-vector contribution ($\hat{W}_{T'}$) from the PV response, since the isoscalar axial coupling of the weak neutral interaction ($\beta_A^{(0)}$) is zero in the Standard Model (at tree level). 
If we further restrict ourselves to elastic scattering by $J^{\pi}=0^+$ nuclear targets, only the Coulomb-type monopole ($C0$) multipole operators contribute to the responses, so that the transverse terms (T) both in the PV and the PC responses do not contribute either. In this situation both responses become trivially proportional \cite{fei75}:
\begin{equation}\label{hadron_ratio}
\frac{W^{PV}}{W^{PC}} = \frac{a_A^e \hat{W}_L}{W_L} = a_A^e\:\beta_V^{(0)}\, ,
\end{equation}
where $\beta_V^{(0)}$ is the vector isoscalar weak neutral coupling of the nucleon and $a_A^e$, as stated above, is the electron weak neutral axial coupling; both can be expressed at tree level in terms of the electromagnetic-weak (Weinberg) mixing angle $\theta_W$ as $\beta_V^{(0)}=-2\sin^2\theta_W$ and $a_V^e=4\sin^2\theta_W-1$, where $\sin^2\theta_W\approx$ 0.23.
The PV asymmetry with the above-mentioned conditions can then be expressed as
\begin{equation}\label{referencevalue}
A = A^0 \equiv -\left[ \frac{G_F \:|Q^2|}{\pi \alpha \sqrt{2}} \right] a_A^e \sin^2\theta_W \cong 3.22 \cdot 10^{-6} \:|Q^2|
\end{equation}
when the momentum transfer is given in fm$^{-1}$.
The nuclear target under consideration in the present work, $^{12}$C, has 0$^+$ angular momentum and parity in its ground state, as well as nominal $T=0$ since $N=Z$. Therefore, a precise measurement of the PV asymmetry allows a precise determination of the values of the Standard Model constants, in particular the mixing angle $\theta_W$, if all of the above mentioned conditions are met, namely, one-boson exchange (no box diagrams), no Coulomb distortion of projectile wave functions, no dispersion effects to non-0$^+$ excited states of $^{12}$C,  absence of strangeness in the nucleons and no isospin-mixing effects. Our current experimental knowledge does not ensure the actual fulfillment of some of these conditions, such as the absence of strangeness in the nucleon \cite{gon13}, and clearly refutes others, such as the isospin purity of the nuclear state which is spoiled by the inescapable Coulomb interaction among protons. Therefore their consequences need to be modeled by theory and extracted from the measured PV asymmetry, introducing theoretical uncertainties in the analysis. Under the term `theoretical uncertainties' we include the variability in the theory describing a given feature, for instance different microscopic nuclear structure models, as well as the experimental uncertainty in the parameters of the models, for instance the strangeness content of the nucleon.

The planned experimental conditions \cite{MITworkshop}, which will be used as starting point and reference case in this study, consist of 150 MeV polarized electrons with a luminosity of $5\cdot 10^{38}$ particles per cm$^2$ per s. Scattered electrons are to be detected with scattering angles between 25$^{\circ}$ and 45$^{\circ}$, typically for a total of 10$^7$ seconds running time (approximately 100 days). An ideal degree of electron polarization of 100$\%$ is assumed here. The statistical uncertainty of the measured PV asymmetry and the model-dependent uncertainty of the corresponding theoretical prediction are to be kept below 0.3$\%$ so that new information on electroweak couplings or on higher-order interaction effects can be extracted.

In addition we explore other possible kinematic regions where theoretical uncertainties and the experimental figure-of-merit are such that the relative error of the asymmetry remains below 0.3$\%$, even when some of the experimental features such the luminosity or the running time are relaxed. We allow for the possibility of an integrated measurement of the PV asymmetry within a given solid angular range, thus reducing the statistical uncertainty. The analysis of different kinematic regions will be based on a set of energies of 150 MeV, 300 MeV and 500 MeV, according to the capabilities of facilities like MESA at Mainz or potentially the FEL accelerator at Jefferson Laboratory.

\begin{figure}
\includegraphics[width=0.5\textwidth]{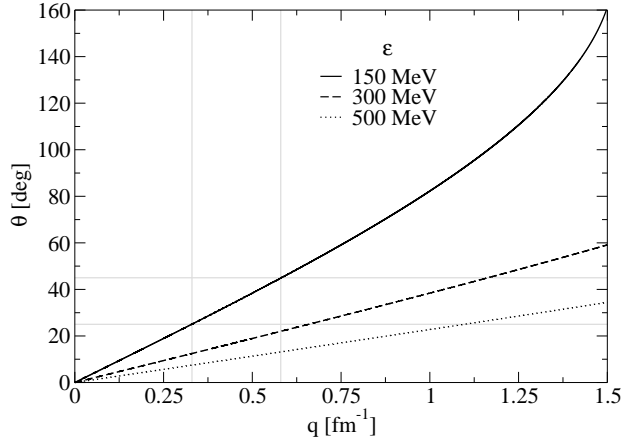}
\caption{Scattering angle as a function of the momentum transfer for different incident energies. The scattering angles of 25$^{\circ}$ and 45$^{\circ}$ are highlighted, as well as the corresponding momentum transfers in our reference case of  $\epsilon=$ 150 MeV for an easy translation of kinematic conditions among the different incident momenta. \label{q_theta}}
\end{figure}

In the processes under study here the incident energy $\epsilon$ of the electrons as well as the scattering angle $\theta$ at which they are detected are under control, both quantities determining the momentum transfer $q$ of the interaction. In the extreme relativistic limit and ignoring the nuclear recoil it is given by
\begin{equation}
q = 2 \:\epsilon \sin(\theta/2) .
\label{kinematics}
\end{equation}
For an easier interpretation of kinematic conditions in terms of momentum transfer or of scattering angle for a fixed incident energy, we show their relationship in Fig. \ref{q_theta} for the incident energies used in this work.

The various effects that will be explored in this work (see below) give rise to different PV
asymmetries which are very hard to distinguish from each other when plotted directly as functions of a kinematic variable. Therefore we plot instead asymmetry deviations for each effect , $\Gamma^X \equiv A^X / A^0 - 1$, defined as the difference between the theoretical PV asymmetry under study, $A^X$ (with only effect $X$ on), and our reference value, $A^0$, divided by the reference value. The total PV asymmetry contains all of the effects and can therefore be written as
\begin{equation}
A^T \approx A^0 \:\left(1+\sum_i \Gamma^{X_i}\right) \:,
\end{equation}
where the interference terms between different effects, of the type $\Gamma^{X_i}\Gamma^{X_j}$, are considered to be small and so are neglected in this analysis. The theoretical uncertainties in any of the PV asymmetry deviations $\Delta \Gamma^{X_i}$ effectively translate into relative theoretical uncertainties of the corresponding PV asymmetries, since
\begin{equation}
\Delta \Gamma^X = \Gamma^{X_a} - \Gamma^{X_b} = \frac{A^{X_a}-A^{X_b}}{A^0} = \frac{\Delta A^X}{A^0} \:.
\end{equation}
For instance, if different reasonable nuclear models yield a
deviation range $\Delta \Gamma^{X_i}=$ 0.01 due to effect $X_i$, the
relative theoretical uncertainty of the PV asymmetry is 1$\%$ (with
respect to $A^0$).

Several sources of theoretical uncertainties are explored in this work. First, the electromagnetic charge of the nuclear target is responsible for the Coulomb distortion of the incoming and outgoing electron wave functions; theoretical uncertainties in the distribution of this charge within the nucleus translate into PV asymmetry uncertainties through the effect of the Coulomb distortion of the projectile wave function, which is computed within the distorted wave Born approximation (DWBA) \cite{ruf82}.  This analysis is carried out using a three-parameter Fermi distribution for the nuclear charge density:
\begin{equation}
\rho(r)=\rho_0 \:\frac{1+\frac{w\:r^2}{c^2}}{1+e^{\frac{r-c}{d}}} \:,
\label{fermi}
\end{equation}
where $c$, $d$, and $w$ are the radius, diffuseness  and central-depression parameters, respectively. Different ground-state charge distributions are then generated by varying the values of these parameters from their fitted values \cite{jag74}, keeping the corresponding rms charge radius within the known experimental range \cite{ang13}. To avoid confusion with isospin-mixing effects (see below), the ground-state neutron distribution is taken to be the same as the charge distribution. 

We study next the effect of isospin mixing in the nuclear ground state due to the electromagnetic interaction, which reveals itself in the form of different proton and neutron distributions. In this case we model the nuclear ground state as an axially-deformed Hartree-Fock mean field employing Skyrme nucleon-nucleon interactions, together with pairing via a BCS approximation \cite{mor09}. Skyrme interactions are effective nucleon-nucleon interactions \cite{sky56,vau}. They include a two-body force in the form of a short-range expansion which leads to momentum dependence ($\vec{k}, \vec{k}'$), and contain the appropriate exchange terms of which only the spin exchange ($P_{\sigma}$) appears explicitly in the final Skyrme expression. An additional term is added to account for two-body spin-orbit interaction. Finally, an extra term introduces a three-body force in the form of a density-dependent ($\rho$) term. A general Skyrme interaction can thus be written as
\begin{eqnarray}\nonumber
V_{12}^{Sk} &=& t_0(1+x_0P_{\sigma})\delta(\vec{r}_1-\vec{r}_2) \:+\: \frac{1}{2} t_1(1+x_1P_{\sigma})[\delta(\vec{r}_1-\vec{r}_2)k^2+k'^2\delta(\vec{r}_1-\vec{r}_2)] \\ \nonumber
&& \:+\: t_2(1-x_2P_{\sigma})\vec{k}'\delta(\vec{r}_1-\vec{r}_2)\vec{k} \:+\:iW_0(\vec{\sigma}_1+\vec{\sigma}_2)\vec{k}' \times \delta(\vec{r}_1-\vec{r}_2)\vec{k} \\
&& \:+\: \frac{1}{6}t_3(1+x_3P_{\sigma})\delta(\vec{r}_1-\vec{r}_2)\:\:\rho^{\alpha}\left(\frac{\vec{r}_1+\vec{r}_2}{2}\right) \:.
\label{skyrme}
\end{eqnarray}
The parameters of the Skyrme interaction that establish the strength of each term ($t_i, x_i, W_0, \alpha$) are fitted to reproduce different properties in different regions of isotopes over the nuclear chart. Although a nucleus as light as $^{12}$C might not be very well suited to such mean-field approaches, we estimate the theoretical variability of the nuclear isospin mixing by using a set of representative Skyrme parametrizations in a Hartree-Fock calculation. A like-nucleon pairing interaction within the BCS approximation is also included, using a fixed pairing energy gap, equal for protons and neutrons, that can be modified to study its effect on the isospin mixing. Another possible indirect source of isospin mixing, an axial deformation in the nuclear ground state, is considered by introducing quadrupole constraints in the Skyrme Hartree-Fock energy functional. Further analysis in the future will require, if available, state-of-the-art ab-initio many-body calculations of the nuclear target ground state. For example, an interesting study could be made using the Green's function Monte Carlo approach \cite{car87}; the $^{12}$C ground state could be obtained using a non-isospin-violating potential where the isospin would be $T=0$ as well as with the AV18 potential which has charge symmetry breaking contributions. And in both cases the Coulomb interaction could be added perturbatively. This would yield estimates of isospin mixing stemming both from the Coulomb interaction as in our present study and also via the charge symmetry breaking terms in the potential. Since the latter are obtained phenomenologically by fixing the potential to fit the systematics of nucleon-nucleon scattering, non-Coulombic isospin-mixing effects would to some extent be incorporated following this procedure.


\section{Results}\label{results}

\begin{figure}
\begin{minipage}[p]{0.48\linewidth}
\centering
\includegraphics[width=0.95\textwidth] {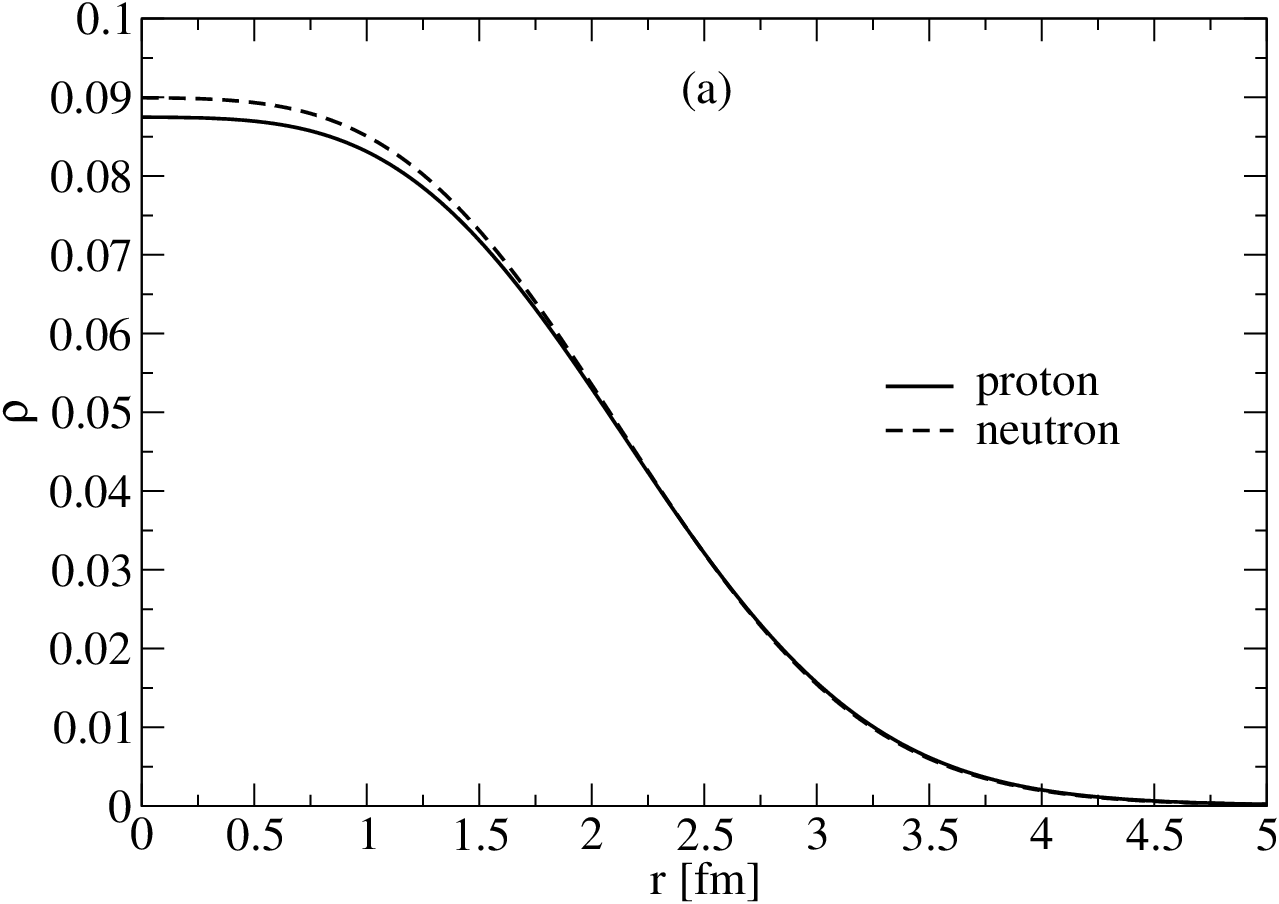}
\end{minipage}
\hspace{0.15in}
\begin{minipage}[p]{0.48\linewidth}
\centering
\includegraphics[width=0.95\textwidth] {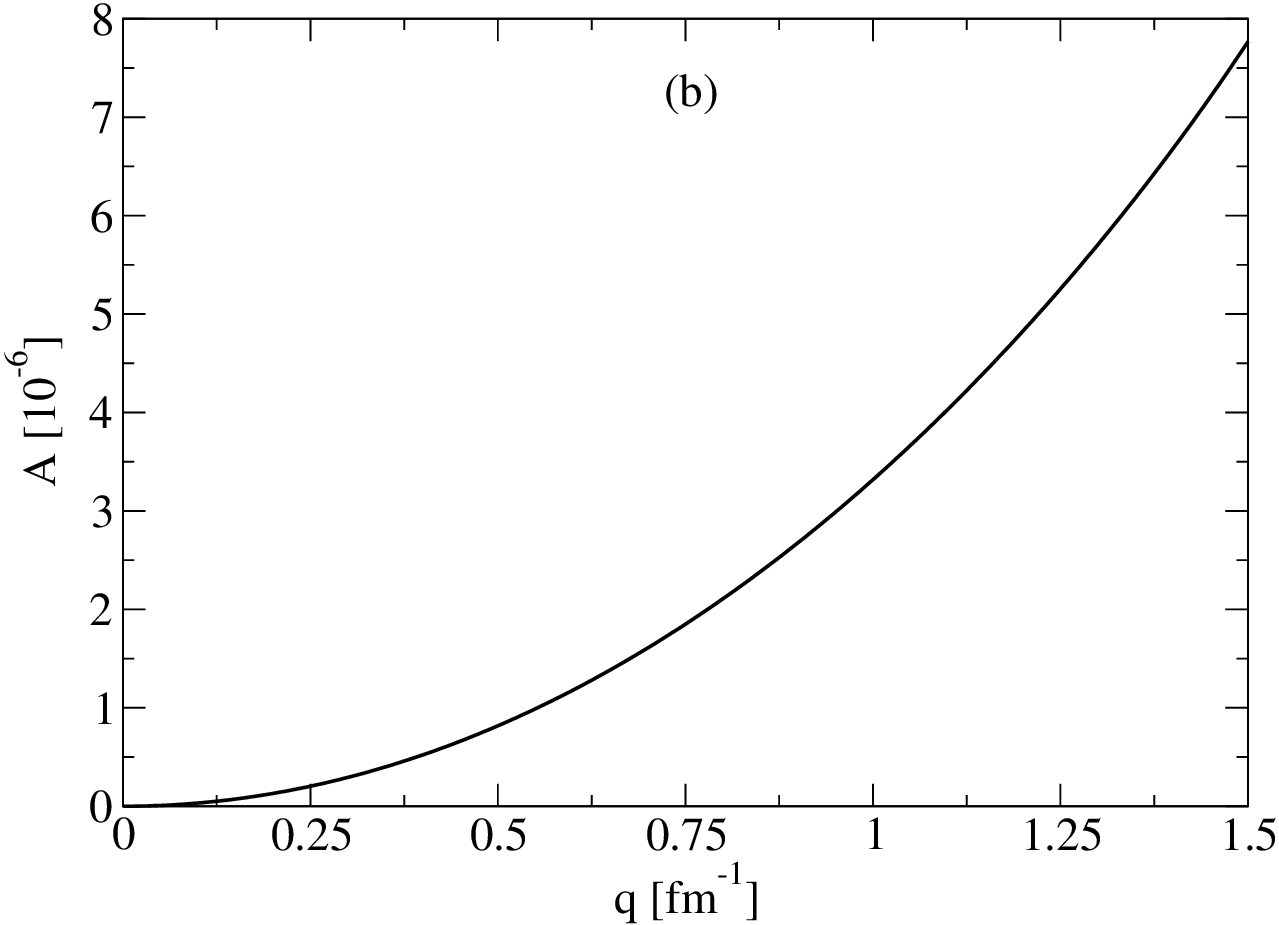}
\end{minipage}
\caption{Left: Proton and neutron radial densities in the ground state of $^{12}$C from a Skyrme (SLy4) Hartree-Fock calculation with BCS pairing for a spherical (self-consistent) shape. Right: PV asymmetry in elastic scattering of polarized electrons with 150 MeV incident energy from a $^{12}$C target, using the ground-state nucleon densities shown on the left. \label{dens_asym_c12}}
\end{figure}

We start by showing in Fig. \ref{dens_asym_c12}, on the left, the proton and neutron densities in the ground state of $^{12}$C and on the right the corresponding PV asymmetry for the scattering of polarized electrons of 150 MeV within DWBA. In both cases the ground-state structure of the nuclear target has been obtained  from a Hartree-Fock calculation using a SLy4 Skyrme interaction \cite{cha98} with BCS pairing for a spherical (self-consistent) nuclear shape.

\begin{figure}
\includegraphics[width=0.5\textwidth] {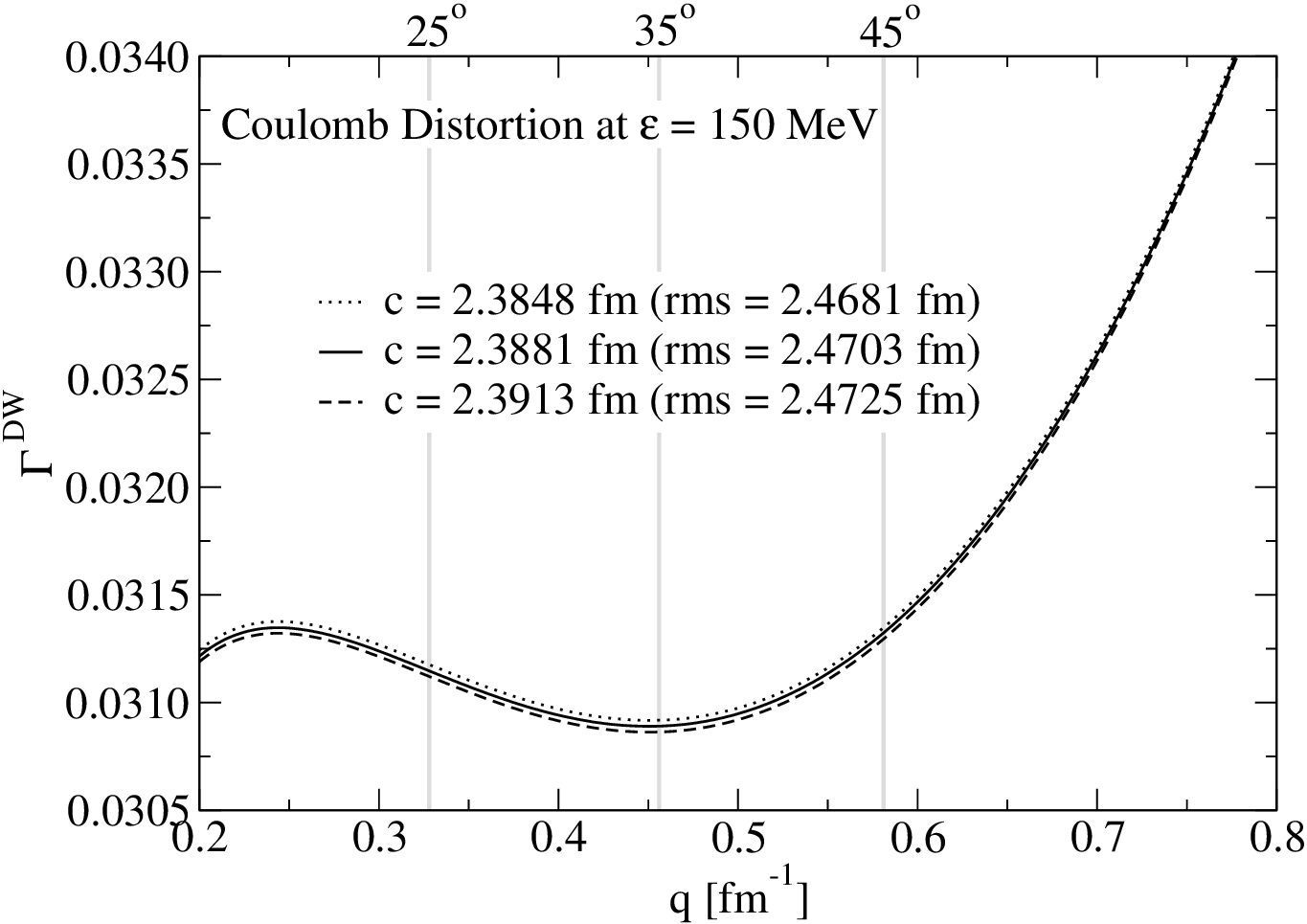}
\caption{PV asymmetry deviation of DWBA results for incident electrons of 150 MeV with respect to those of PWBA, as a function of the 3-momentum transfer $q$ (lower axis) or scattering  angle (upper axis). Three results are shown using different values of the radius parameter $c$ of the Fermi charge distribution (compatible with the uncertainties in the experimental rms charge radius). \label{dev_dist_e150}}
\end{figure}

\begin{figure}
\begin{minipage}[p]{0.48\linewidth}
\centering
\includegraphics[width=0.95\textwidth] {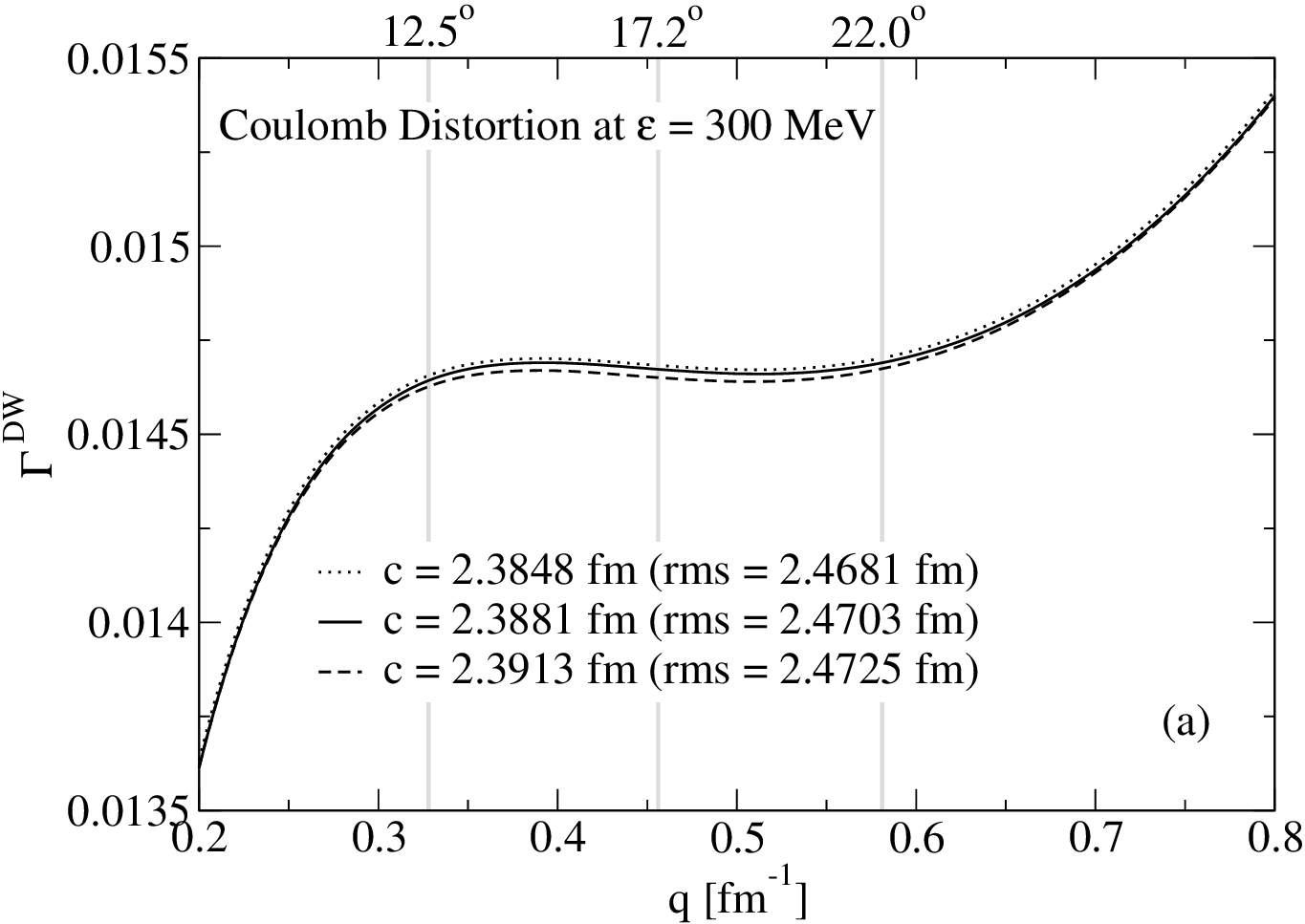}
\end{minipage}
\hspace{0.15in}
\begin{minipage}[p]{0.48\linewidth}
\centering
\includegraphics[width=0.95\textwidth] {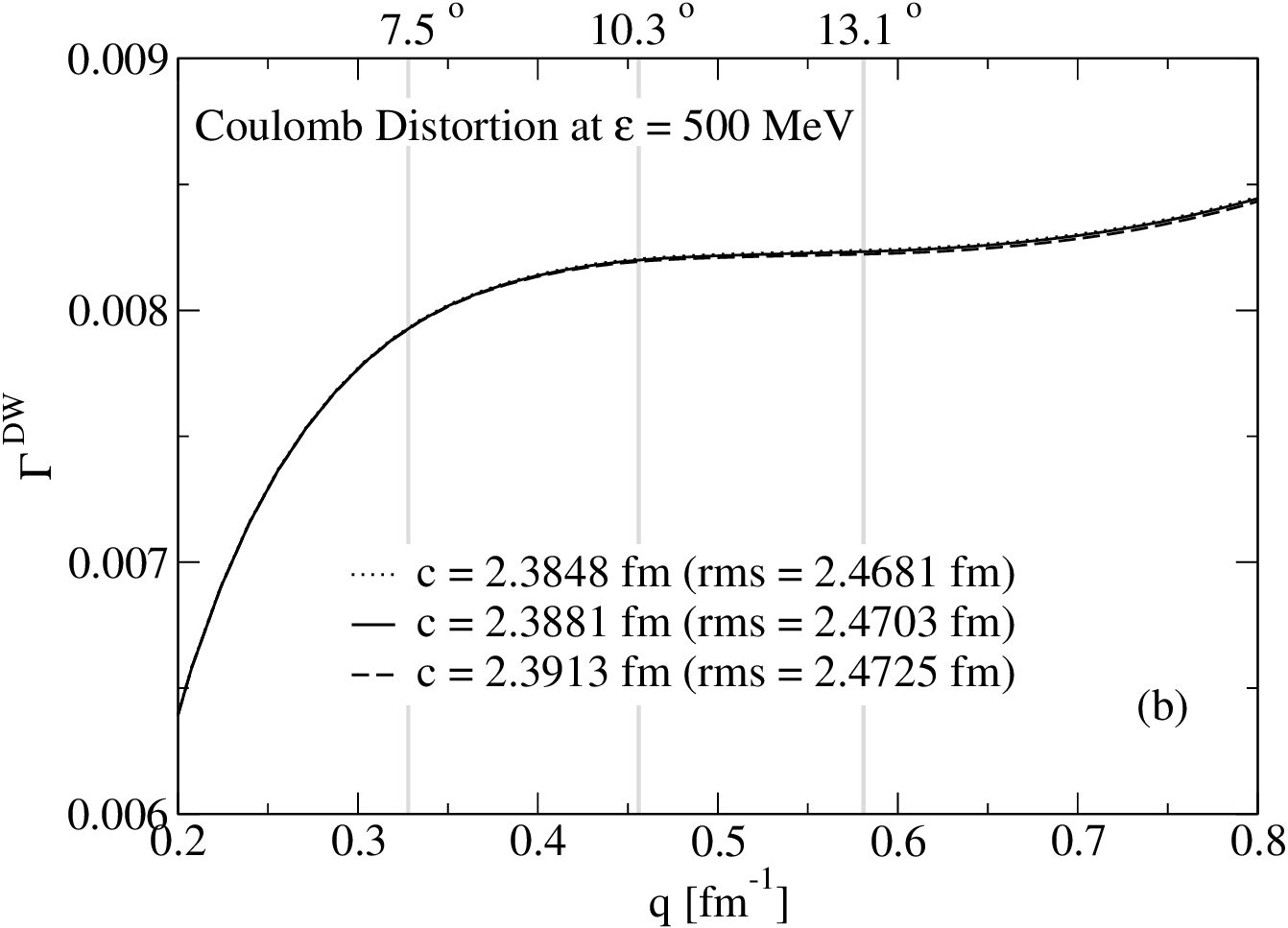}
\end{minipage}
\caption{Same as for Fig. \ref{dev_dist_e150} but now for higher energies of the incident electrons. Left:  For 300 MeV incident electrons. Right: For 500 MeV incident electrons. \label{dev_dist_higher_energies}}
\end{figure}

We analyze first the effect of the nuclear charge distribution uncertainty on the PV asymmetry. 
Figs. \ref{dev_dist_e150} and  \ref{dev_dist_higher_energies} show the PV asymmetry deviation due to Coulomb distortion effects, {\it i.e.} the deviation of the DWBA asymmetry with respect to the PWBA calculation, $\Gamma^{DW} = A^{DW} / A^0 - 1$. Different charge distributions (equal to the neutron distributions) have been used in the ground state of $^{12}$C, obtained by varying the radius parameter $c$ of the Fermi distribution in Eq. (\ref{fermi}). In the kinematic range of interest for 150 MeV incident electrons the effect of Coulomb distortion of the electron wave function lies around 3$\%$ with a theoretical spread of 0.01$\%$, as shown in Fig. \ref{dev_dist_e150}; for the higher incident energies shown in Fig. \ref{dev_dist_higher_energies} (300 MeV and 500 MeV), the Coulomb distortion effects are smaller, as are the theoretical spreads. In all cases these theoretical spreads are well below the desired limit of 0.3$\%$. The same analysis has been performed by varying the diffuseness and the central-depression parameters of the Fermi distribution, keeping the charge rms radius within the experimental range. The same conclusions as with the radius parameter variation can be drawn.

\begin{figure}
\begin{minipage}[c]{0.48\textwidth}
\centering
\includegraphics[width=0.95\textwidth] {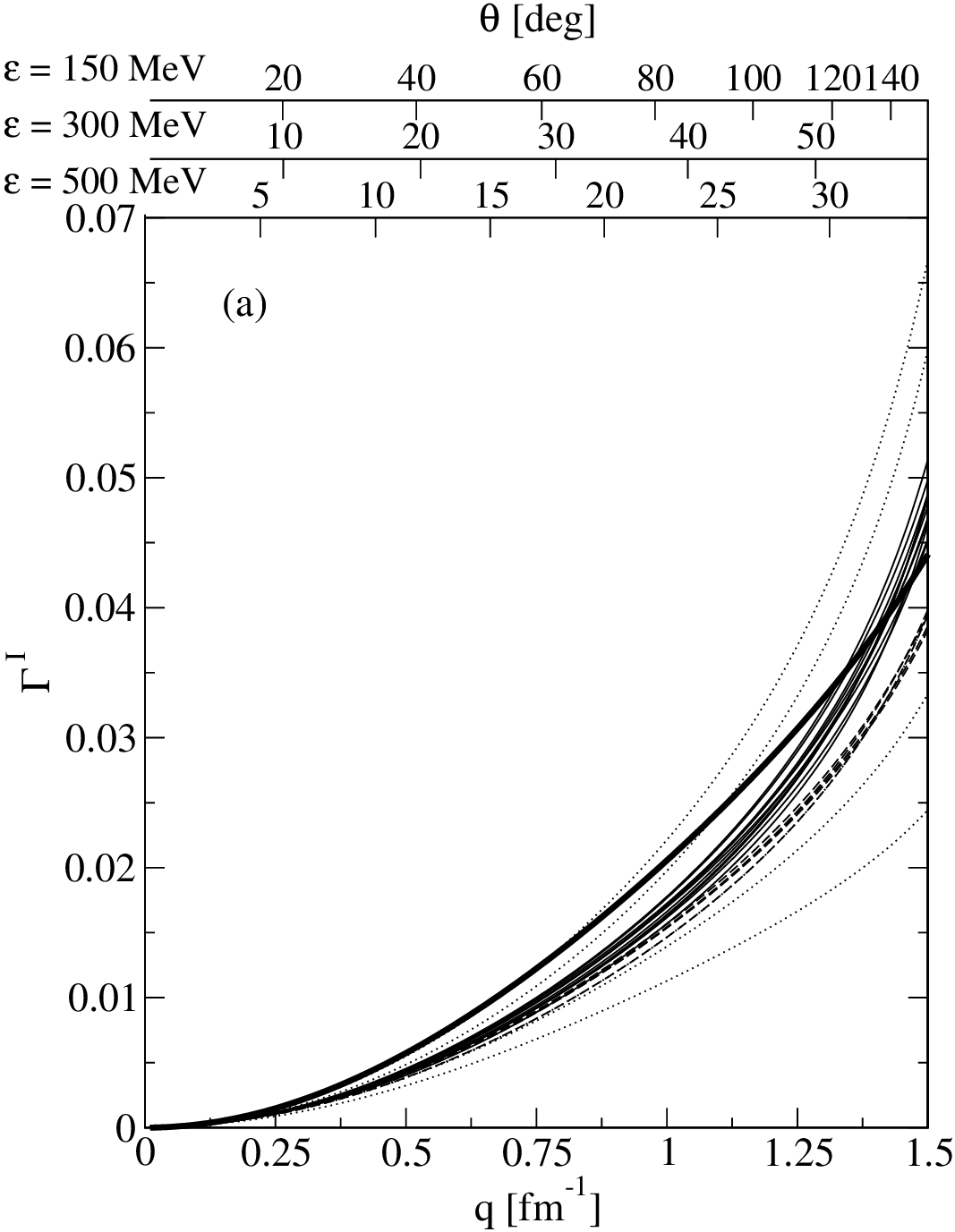}
\end{minipage}
\hspace{0.1in}
\begin{minipage}[c]{0.48\textwidth}
\centering
\includegraphics[width=0.95\textwidth] {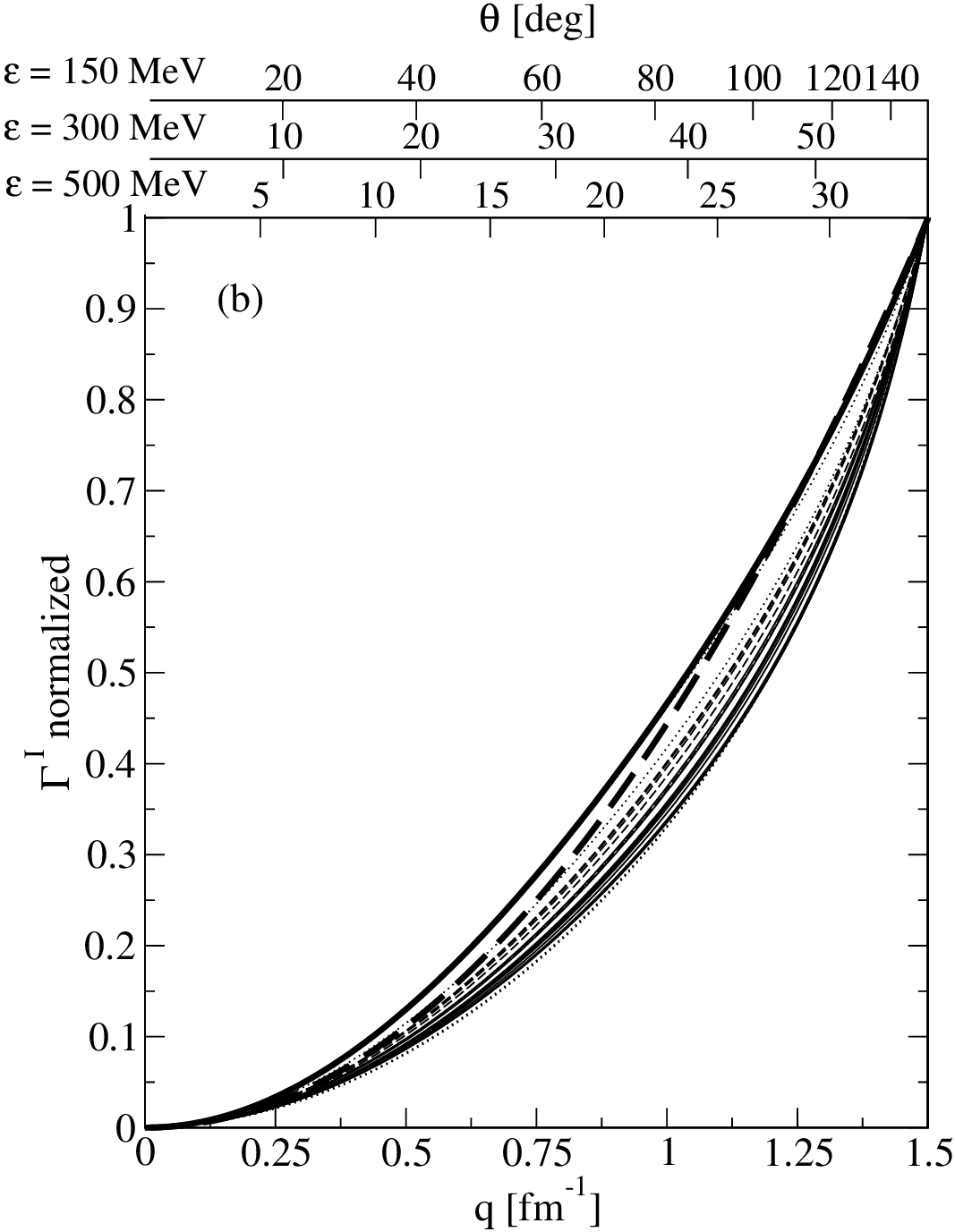}
\end{minipage}
\caption{Left: Deviation of the PV asymmetry having isospin mixing with respect to the value in the absence of mixing, as a function of the momentum transfer $q$ in the lower axis and indicating the corresponding scattering angles in the upper axes for three incident energies, 150 MeV, 300 MeV and 500 MeV. Several results are shown for different Skyrme forces used in a Hartree-Fock calculation (thin solid and dashed lines for two groups of similar results, and thin dotted lines for outliers), together with a relativistic mean field calculation using a NLSH lagrangian parametrization (thick solid line). Right: Same as for figure on the left but with all the curves normalized to 1 at q=1.5 fm$^{-1}$. Thick dashed line shows a pure $q^2$ dependence for comparison.  \label{dev_dist_skyrme}}
\end{figure}

Another contribution to the PV asymmetry comes from the nuclear isospin mixing
in the form of different proton and neutron distributions, which we estimate using a Skyrme HF mean field calculation with BCS pairing, as discussed above. The isospin mixing in this model results exclusively from the Coulomb interaction between protons. The relevant PV asymmetry deviation in this case compares
the asymmetry where isospin mixing is present with the one where the ground state has zero isospin, $\Gamma^I = A^I / A^0-1$.
In Fig.~\ref{dev_dist_skyrme} we plot results with several Skyrme forces (see \cite{dut12} and references therein) to show both the average size of the isospin mixing effect as well as an estimation of the theoretical uncertainty within the model. The overall value of this deviation is seen to be 0.3--0.5$\%$ in the region of interest,  although most of the Skyrme forces tested here yield deviations lying within an even smaller range. Some of the outliers in the subset used in this work can be questioned on the grounds that in some cases not all of the parameters in the interaction were adjusted but were fixed, or in others that some of the relevant terms in the interaction were absent. It should also be noted that the various interactions were obtained by emphasizing good agreement for specific properties (binding energies, energy levels, the nature of the deformation, rms radii, BE2s), but not necessarily all properties simultaneously, and it is not obvious what matters most for the analysis of the PV asymmetry. Furthermore, as stated above, the interactions being used were not specifically designed for $^{12}$C, and one sees reflections of this in the fact that some are better than others at reproducing the position of the diffraction minimum in the elastic cross section. Similar conclusions could be expected from the whole set of near 250 different Skyrme parametrizations that can be found in the literature and that work reasonably well in describing a given set of properties of finite nuclei or of nuclear matter. Concerning the latter, a comprehensive analysis of the Skyrme interactions has been performed in \cite{dut12} in relation to nuclear matter constraints to find that only a few of them pass the test. However, we find it risky to endorse or rule out some Skyrme parametrizations for our finite (actually light) $N=Z$ nucleus based only on nuclear matter evaluations, even those related to the symmetry energy which might seem in principle to be particularly sensitive to proton versus neutron distributions and therefore to isospin-mixing effects.
To a very good approximation, the PV asymmetry deviation due to isospin mixing shows approximately a quadratic $q$ dependence, especially in the low momentum transfer region. This behavior can be easily traced back to the momentum transfer dependence of the Coulomb monopole operator involved in the PV asymmetry under study in this work.
Together with the Skyrme Hartree-Fock results, we show in Fig.~\ref{dev_dist_skyrme} the isospin-mixing deviation from a relativistic mean field approach, a Dirac-Hartree study performed starting with the NLSH parametrization of the lagrangian density \cite{rein86}. The resulting deviation lies within the range defined by the set of Skyrme forces described above.

\begin{figure}[t]
\begin{minipage}[p]{0.48\linewidth}
\centering
\includegraphics[width=0.95\textwidth] {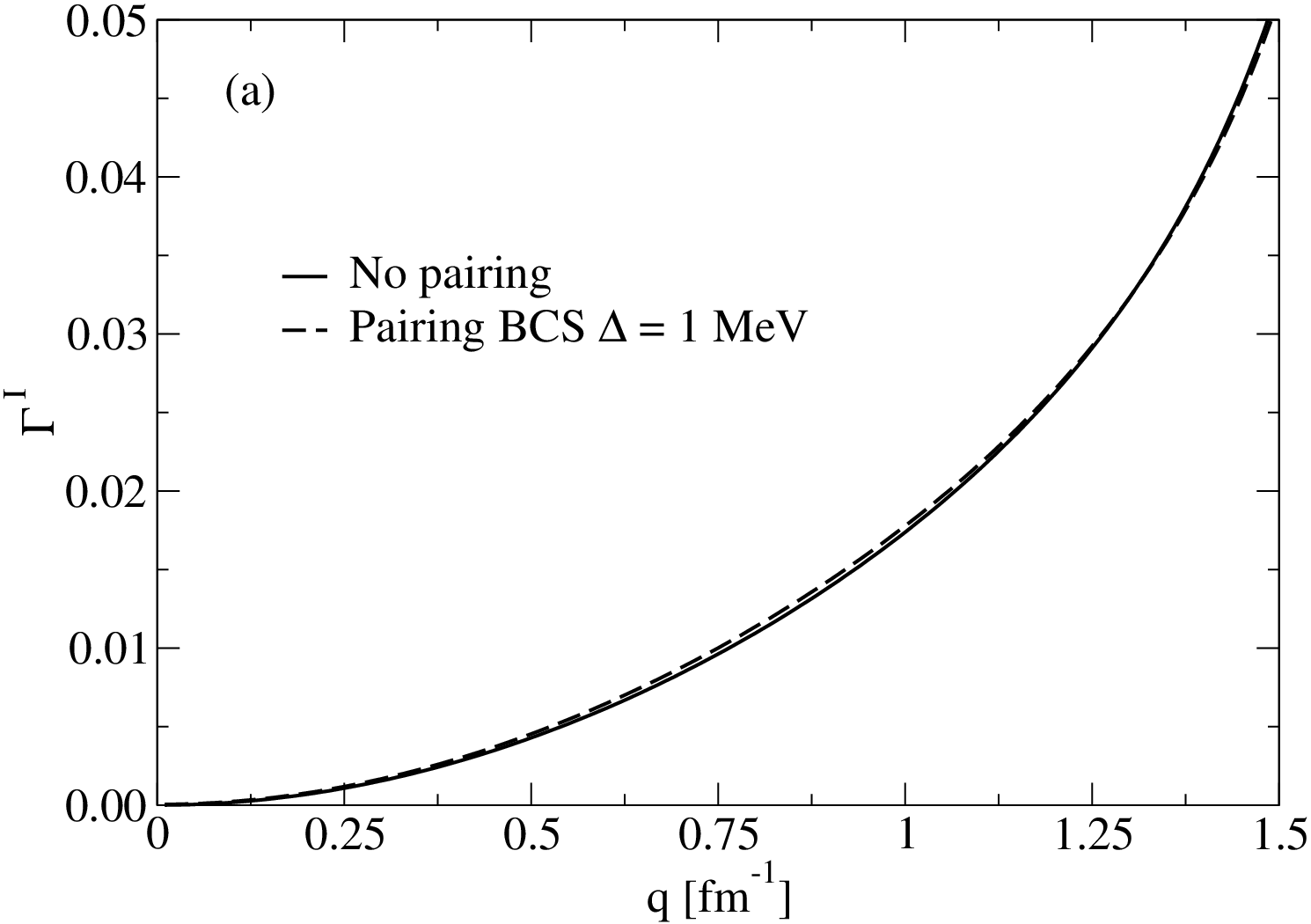}
\end{minipage}
\hspace{0.15in}
\begin{minipage}[p]{0.48\linewidth}
\centering
\includegraphics[width=0.95\textwidth] {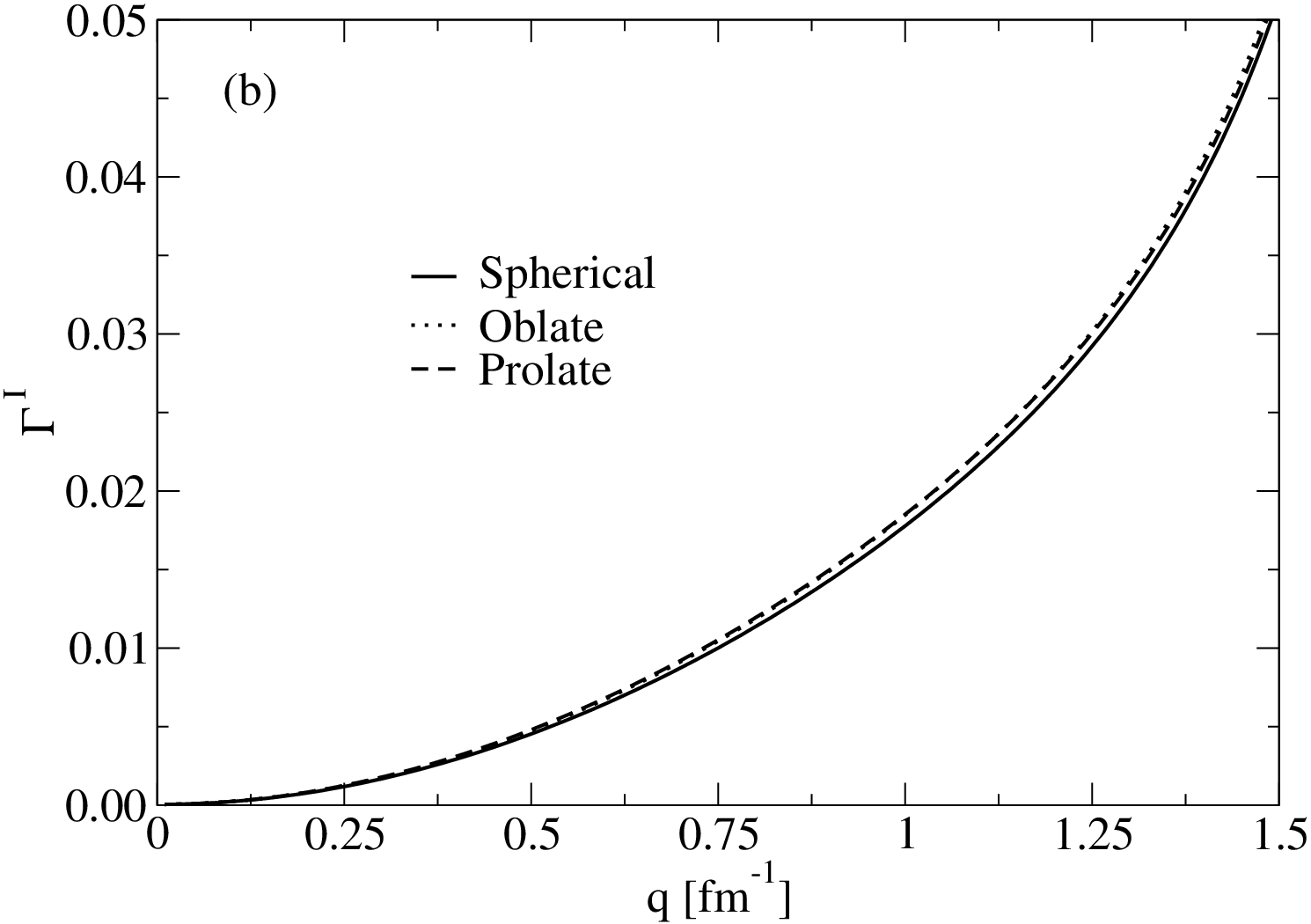}
\end{minipage}
\caption{Left: Deviation of the PV asymmetry having isospin mixing with respect to the value
  in the absence of mixing, as a function of the momentum transfer $q$, for two different Skyrme Hartree-Fock  calculations, with and without residual pairing interactions between like-nucleons within BCS approximation. Right: The same, but now for three different axially-symmetric nuclear shapes from a deformed Skyrme Hartree-Fock  calculation, spherical, oblate and prolate. \label{pairing_defor_c12}}
\end{figure}

Other modifications that can be considered in the nuclear mean field with a potentially different impact on the proton and neutron distributions, {\it i.e.} contributing to the isospin mixing, are shown in Fig. \ref{pairing_defor_c12}. On the left we show the effect of including or removing a residual pairing interaction between like-nucleons in the Skyrme Hartree-Fock calculation. The strength of this interaction is introduced through a fixed pairing energy gap within the BCS approximation, whose value is difficult to estimate and is therefore an actual source of uncertainty. On the right we plot the isospin-mixing deviations for different axially symmetric shapes of the mean field (spherical, oblate and prolate), which are computed using a quadrupole constraint in the Hartree-Fock calculation. This is another source of uncertainty since the ground-state shape of $^{12}$C is not very well known (see the discussion in \cite{mor09}) and could even imply triaxial deformations that we nevertheless model through axial deformations. The former could be related to a three-alpha cluster structure interpretation \cite{fyn05}, which can be revealed in calculations beyond mean field involving more than just one Slater determinant to account for deformation and clustering \cite{fuk13}.
Notwithstanding the above discussion, as can be seen in the figures, all the variations considered, namely different Skyrme parametrizations, pairing strengths and axially symmetric shapes, yield a theoretical spread lower than 0.1$\%$, which is below the target goal of a few tenths of a percent.

We now turn to a brief discussion of the situation where the experimental
resolution is not sufficient to resolve the ground state and accordingly
where the PV asymmetry arises from a sum over several excited states
together with the ground state. Let us assume that the sum runs over $i=0\ldots n$, with $0,1,\ldots$ denoting the ground state, first excited
state, \textit{etc.} The total asymmetry is given by
\begin{equation}
A=\sum_{i=0}^{n}f_{i}A_{i}\:, \quad\qquad \text{where} \quad\qquad f_{i}\equiv \frac{\sigma _{i}}{\sum_{j=0}^{n}\sigma _{j}}\:,
\end{equation}
with $A_{i}$ being the PV asymmetry for excitation of the $i^{th}$ state and 
$\sigma _{i}$ being the (parity-conserving) cross section for a transition
from the ground state to the $i^{th}$ excited state and one has $\sum_{i=0}^{n}f_{i}=1$. Defining the deviation from the reference asymmetry
as above one then has
\begin{eqnarray}
\Gamma ^{inel} &\equiv & A/A^{0}-1 = \left( f_{0}-1\right) + \sum_{i=1}^{n}f_{i} \:(A_{i}/A^{0}) .
\end{eqnarray}
Since the elastic scattering cross section, namely the contribution from the
Coulomb monopole charge form factor, is proportional to $Z^{2}$, whereas the
inelastic cross sections are not coherent, one expects at least to have $f_{i}\sim 1/Z^{2}$ for $i\geq 1$ and thus that $f_{0}-1\sim -n/Z^{2}$. In
fact, at low momentum transfers the inelastic multipole matrix elements are
further suppressed by powers of $q/q_{N}$ where $q_{N}$ is a characteristic
nuclear momentum transfer scale, roughly 1 fm$^{-1}$. Thus, even from such
rough arguments one expects a minor contribution from inelastic transitions,
unless the resolution is so poor that a very large range of energies must
occur in the sums above.

In the specific case of $^{12}$C, first assuming that the energy resolution
is sufficient to involve only the $T=0$ excited states (\textit{i.e.,}
better than 15.11 MeV; here we do not consider isospin mixing in the excited
states, although in a full analysis that can be taken into account),
assuming no strangeness contributions and working at tree level in the
Standard Model (where the isoscalar axial-vector coupling is zero) one has a
very simple answer, namely $\Gamma ^{inel}=0$, since all weak neutral
current multipole matrix elements are proportional to the corresponding electromagnetic matrix elements with a universal coupling. Thus any non-zero result must come from having
strangeness or from taking into account beyond-tree-level contributions to
the isoscalar axial-vector matrix elements. The former occurs because for
inelastic contributions in general both $G_{E}^{(s)}$ and $G_{M}^{(s)}$ can
occur, in contrast to the elastic scattering result where
magnetic strangeness enters only as a very small relativistic spin-orbit
contribution (see the following discussion). The latter implies that the VA interference response must be
taken into account; however, because of the smallness of the vector leptonic
coupling and because this contribution is suppressed at the forward
scattering angles being considered in this work, one expects a very small
contribution here.

The situation is even clearer if the experimental resolution is good enough
to require one to take into account only the first two excited states of
carbon, the $2^{+}$ state at 4.4389 MeV and the $0^{+}$ state at 7.6542 MeV \cite{fir99}.
A transition to the latter has the same characteristics as the elastic case,
namely with the same proportionality of the WNC and EM matrix elements, and
hence has the same asymmetry, $A_{2}=A_{0}$. The former can be more
complicated in that both C2 and E2 multipoles enter and the E2 has both
convection and magnetization current contributions, convection involving $G_{E}^{(s)}$ and magnetization $G_{M}^{(s)}$, while the C2 involves only $G_{E}^{(s)}$. In fact, at the low momentum transfers of interest in this work the E2 is known to be dominated by the convection current \cite{flanz} and thus one finds that $A_{1}\simeq A_{0}$ as well. Indeed, putting in
estimates for the form factors involved and taking into account strangeness
content as discussed below, one finds that $\Gamma ^{inel}$ is below the 10$^{-3}$ level and hence inconsequential for the present discussions.

Regarding the possible effects of meson-exchange currents (MEC), we note that MEC effects of non-strange type will cancel in a situation where there is no isospin mixing and no strangeness. If isospin mixing is present, but no strangeness, isoscalar and isovector matrix elements are modified in different ways, typically at the
10$\%$ level or less at low momentum transfers \cite{dub76}, and so the results above may change by roughly this amount leading to uncertainties of typically a few parts in 1000. Finally, with isospin mixing and strangeness in the nucleons (one-body strangeness) the dominant effects are from the latter; two-body strangeness effects at low $q$ should be very small according to previous studies \cite{msd}.

\begin{figure}
\includegraphics[width=0.5\textwidth]{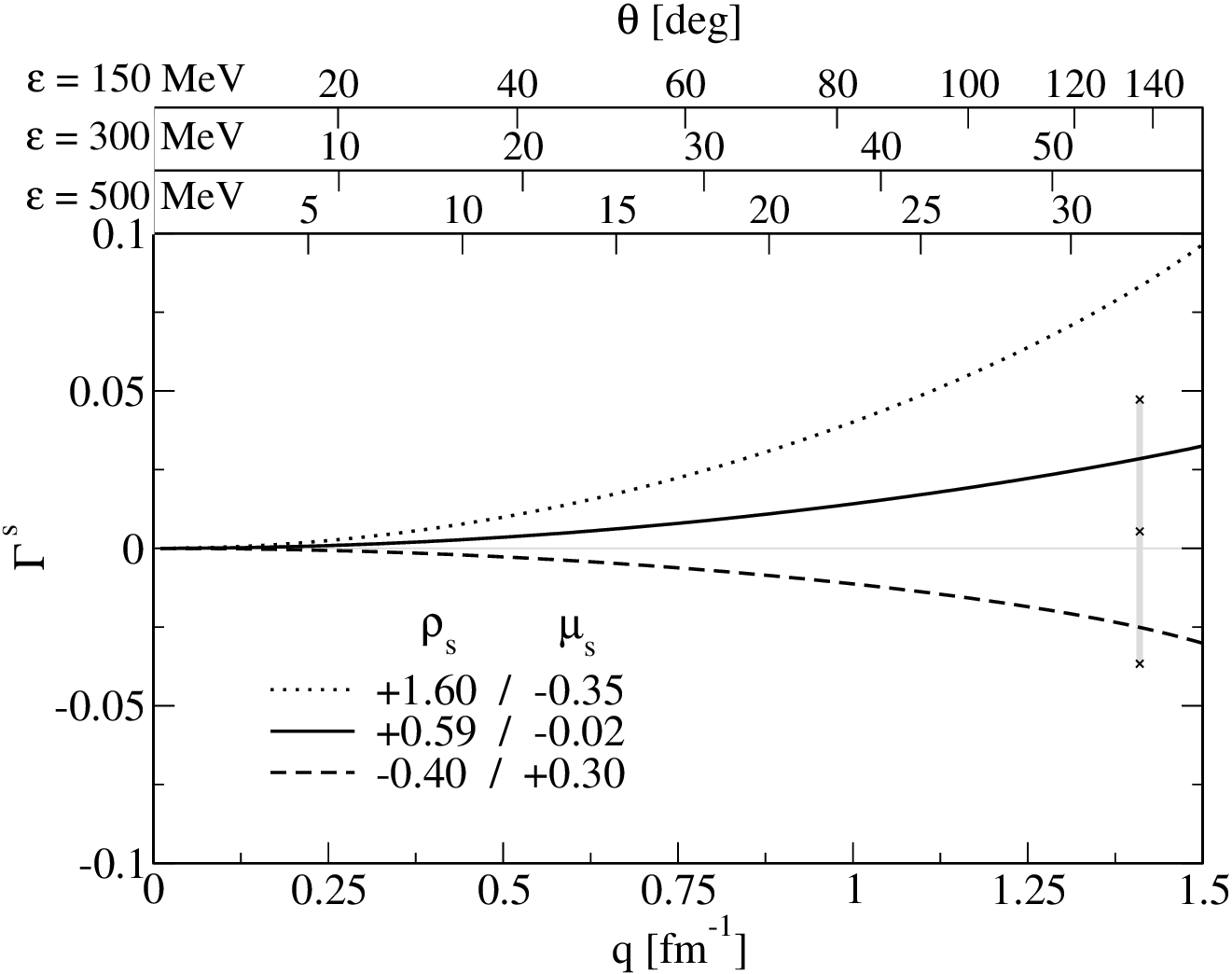}
\caption{Deviation of the PV asymmetry due to the strangeness content in the nucleon with respect to that without strangeness, as a function of the momentum transfer $q$ in the lower axis and of the scattering angle in the upper axes for three incident energies, 150 MeV, 300 MeV and 500 MeV. Three results are shown for the limiting and central combined values of the experimental range of the electric $\rho_s$ and magnetic $\mu_s$ nucleon strangeness content parameters  \cite{mus93,mus94,gon13}. The experimental range extracted from the HAPPEX-He experiment is also shown (thick grey line) \cite{ach07}. \label{dev_strange}}
\end{figure}

We focus finally on the uncertainties stemming from the electric and magnetic strangeness form
factors of the nucleon. The relevant asymmetry deviation in this case is $\Gamma^s = A^s / A^0 - 1$, where $A^0$ is the standard asymmetry (with no strangeness). The electric and the magnetic strangeness form factors of the nucleon can be parametrized as
\begin{equation}
G_E^{(s)}=\frac{\rho_s\tau}{(1+4.97\tau)^2}
\label{strange_elec_ff}
\end{equation}
\begin{equation}
G_M^{(s)}=\frac{\mu_s}{(1+4.97\tau)^2}\, ,
\label{strange_magn_ff}
\end{equation}
where $\tau=|Q^2|/4m^2_N$ and $\rho_s$ and $\mu_s$
are the electric and magnetic strangeness content parameters respectively \cite{mus93,mus94,gon13}. The current experimental values of
these parameters are interrelated and range from $\rho_s=$ 1.60 and
$\mu_s=$ -0.35 to $\rho_s=$ -0.40 and $\mu_s=$ 0.30, with a central
value of $\rho_s=$ 0.59 and $\mu_s=$ -0.02 \cite{gon13}. The
asymmetry deviations corresponding to these limiting and central
values are shown in Fig.~\ref{dev_strange}, yielding an overall
effect from 0.2 to 1$\%$ in the region of interest. The spread
arising from these three cases lies between 0.5 and 1.5$\%$.  The
most relevant contribution to these effects comes from the electric
strangeness, since magnetic strangeness in the monopole matrix
element arises only through a rather suppressed spin-orbit
correction of relativistic origin \cite{mor09}. However, a larger effect of the magnetic strangeness may appear in inelastic transitions, as discussed above.

It is clear from the previous discussion that current experimental uncertainty on the strangeness content of the nucleon is a critical source of uncertainty in the PV asymmetry. In order to reduce this uncertainty below the critical value, $\Delta \Gamma^{s} < 0.003$ in our case, one should, according to Fig. \ref{dev_strange}, restrict the measurements to a kinematic region where the momentum transfer $q$ is lower than a given value $q_0$, which corresponds to a given value of the scattering angle $\theta_0$ different for each incident energy $\epsilon$. At the same time the kinematic region of measurement should provide a large enough number of events so that the statistical error lies below the critical value, $(\Delta A / A)_{exp} < 0.003$. The latter is clearly the experimental contribution to the asymmetry uncertainty, whereas the former, $\Delta \Gamma$, is the theoretical contribution; systematic errors are not dealt with in this study.

To increase the number of PV events detected, and therefore to reduce the statistical uncertainty of the asymmetry, we consider the possibility of detection within a wide solid angle of a given polar coverage and a fixed 2$\pi$ azimuthal angle coverage without segmentation, {\it i.e.} without angular bins. In this situation the asymmetry may not be considered constant within the solid angle of detection and the polar angle dependence must be taken into account.

According to the definition of the PV asymmetry, Eq. (\ref{asymmetry_sigmas}), the difference between the number of electrons with opposite spin projections detected at a given scattering angle is $N ^+(\theta) - N^-(\theta) = A(\theta) \:N_T(\theta)$, where $N_T$ stands for the total number of events ($N^+ + N^-$). The total PV asymmetry after integration over scattering (polar) angle can then be written as
\begin{equation}
A = \frac{N^+ - N^-}{N_T} = \frac{\int^{\theta_f}_{\theta_i} \:d\theta \:[N^+(\theta) - N^-(\theta)]}{\int^{\theta_f}_{\theta_i} \:d\theta \:N_T(\theta)} = \frac{\int^{\theta_f}_{\theta_i} \:d\theta \:A(\theta) \:N_T(\theta)}{\int^{\theta_f}_{\theta_i} \:d\theta \:N_T(\theta)}.
\end{equation}
On the other hand the statistical uncertainty of the asymmetry is
\begin{equation}
\Delta A = N_T^{-1/2} = \left[\int^{\theta_f}_{\theta_i} \:d\theta \:N_T(\theta)\right]^{-1/2}.
\end{equation}
From the previous two equations the relative statistical uncertainty of the PV asymmetry can be written as
\begin{equation}
\frac{\Delta A}{A} = \frac{\left[\int^{\theta_f}_{\theta_i} \:d\theta \:N_T(\theta)\right]^{1/2}}{\int^{\theta_f}_{\theta_i} \:d\theta \:A(\theta) \:N_T(\theta)} =  \frac{1}{\left[\Delta\phi \;\;L \;\;T \right]^{1/2}} \;\;\frac{\left[\int^{\theta_f}_{\theta_i} \:d\theta \; \frac{d\sigma}{d\Omega}(\theta) \;\sin\theta\right]^{1/2}}{\int^{\theta_f}_{\theta_i} \:d\theta \;A(\theta) \;\frac{d\sigma}{d\Omega}(\theta) \;\sin\theta} \, ,
\end{equation}
where the total number of events has been replaced by the expression
\begin{equation}
N_T(\theta) = \frac{d\sigma}{d\Omega}(\theta) \:\Delta\phi \:\sin\theta \:L \:T \, ,
\end{equation}
with $d\sigma/d\Omega$ the differential cross section with respect to the solid angle of detection, $\Delta\phi$ the azimuthal angular coverage, $L$ the luminosity of the incident beam and $T$ the running time of the experiment. This leads to
\begin{equation}
\frac{\Delta A}{A} = \frac{1}{\left[\Delta\phi \;\;L \;\;T \right]^{1/2}} \;\;\frac{\left[\int^{\theta_f}_{\theta_i} \:d\theta \; \frac{d\sigma}{d\Omega}(\theta) \;\sin\theta\right]^{1/2}}{\int^{\theta_f}_{\theta_i} \:d\theta \;A(\theta) \;\frac{d\sigma}{d\Omega}(\theta) \;\sin\theta}.
\end{equation}

\begin{figure}
\includegraphics[width=0.5\textwidth]{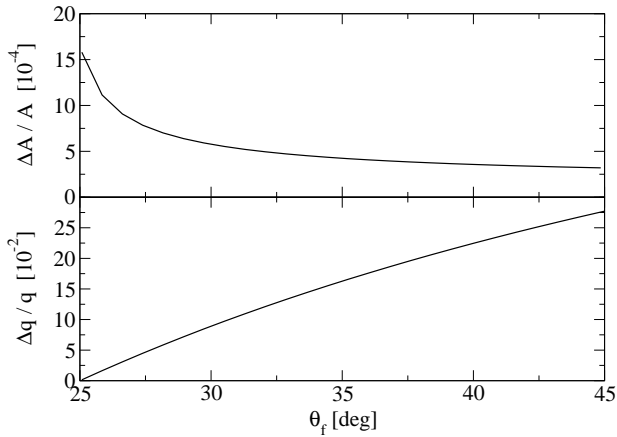}
\caption{Upper panel: Relative statistical uncertainty of the PV asymmetry for 150 MeV polarized electrons given luminosity $5^.10^{38}$ s$^{-1}$cm$^{-2}$ and 10$^7$ s running time, as a function of the final polar angle $\theta_f$ of the detector starting at $\theta_i=$ 25$^{\circ}$, with 360$^{\circ}$ azimuthal coverage. Lower panel: Relative uncertainty of momentum transfer  corresponding to this solid angle.\label{rel_error_vs_th}}
\end{figure}

The relative statistical error of the asymmetry decreases as the polar angle coverage of the detector increases, but at the same time the uncertainty in the dependent variable (the scattering angle or equivalently the momentum transfer), increases. Both uncertainties are shown in Fig. \ref{rel_error_vs_th}  as a function of the final polar angle of the detector, using $\theta_i = 25^{\circ}$ as initial angle and with 360$^{\circ}$ azimuthal angular coverage.

\begin{figure}
\includegraphics[width=0.5\textwidth]{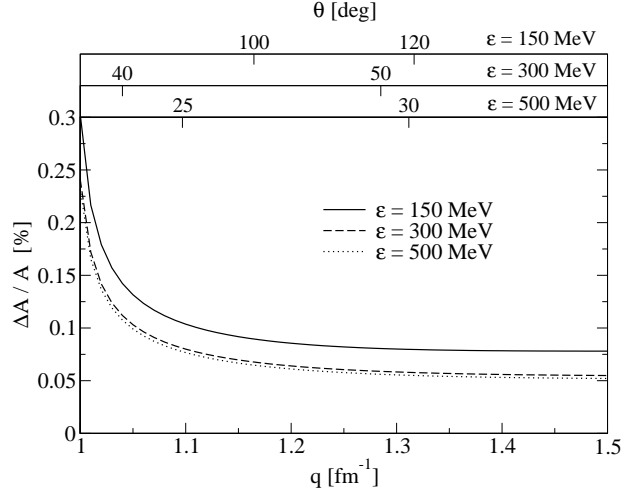}
\caption{Relative statistical uncertainty of the PV asymmetry for polarized electrons of different energies given luminosity 5$\cdot$10$^{38}$ s$^{-1}$cm$^{-2}$ and 10$^7$ s running time, as a function of the momentum transfer in the range 1 - 1.5 fm$^{-1}$ in the lower axis and the corresponding polar (scattering) angles for different incident energies in the upper axes. \label{energies_err_rel_vs_q}}
\end{figure}

\begin{figure}
\vspace{10pt}
\includegraphics[width=0.5\textwidth]{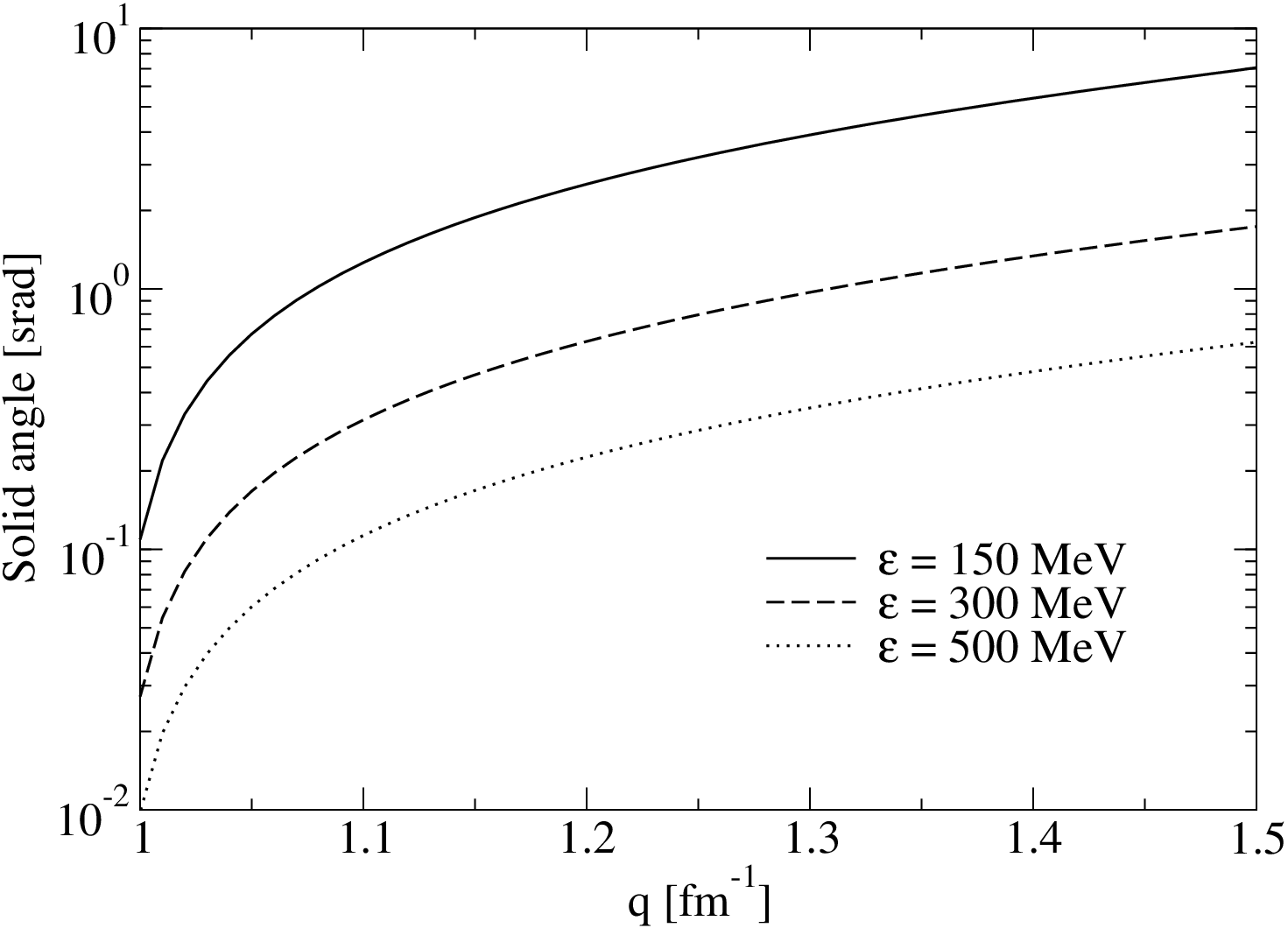}
\caption{Solid angle of scattering corresponding to a momentum transfer range 1 - 1.5 fm$^{-1}$ for different incident energies. \label{energies_solidangle_vs_q}}
\end{figure}

In order to reduce experimentally the uncertainty in the nucleon strangeness content it would seem convenient to focus on a kinematic region where this uncertainty is large and where at the same time those corresponding to other sources remain sufficiently small. The $q^2$ dependence of the strangeness contribution suggests focusing on a momentum transfer region between 1 and 1.5 fm$^{-1}$ where, as can be seen in Fig. \ref{dev_strange}, the uncertainty in the strangeness contribution is large and at the same time the minima of the nuclear form factors do not yet play a role (see \cite{mor09}). In Fig. \ref{energies_err_rel_vs_q} we show the relative error of the PV asymmetry in this region as a function of the final polar angle of the detector or the corresponding momentum transfer, the initial one being $q_i$ = 1 fm$^{-1}$; it can be seen there that larger incident energies result in smaller relative errors even if the solid angle of detection is also smaller, as shown in Fig. \ref{energies_solidangle_vs_q}.
Assuming a small contribution from any other effect, a precise measurement of the PV asymmetry in this region would reduce the strangeness content uncertainty, which is then translatable to any other kinematic region, in particular to low $q$, according to Eqs. (\ref{strange_elec_ff}) and (\ref{strange_magn_ff}). For instance, a 2$\%$ precision in a measurement of the asymmetry at $q=$ 1.5 fm$^{-1}$ would reduce to around 0.2$\%$ at $q=$ 0.5 fm$^{-1}$.
However, the previous strategy would not be sufficient by itself if the size and the uncertainty of another effect play a relevant role in the same kinematic region, as seems to be the case with the isospin-mixing discussed above. Several measurements at different momentum transfers would be helpful to distinguish two different contributions if different dependences on this kinematic variable were expected for them. It is not the case for the two effects under analysis now, since the isospin-mixing contribution also shows a $q^2$-dependence to a very good approximation. Thus we are left with the following situation: both the effects from isospin-mixing (at least as evaluated in this study) and from electric strangeness at relatively low momentum transfers in PV scattering from $^{12}$C track proportionally to $q^2$. This is both good and bad. On the one hand, it could be considered to be good if the primary goal is to determine the Standard Model/radiative effects at very low momentum transfers, for then the isospin/strangeness contributions taken together could be determined at somewhat higher momentum transfers (between 1 and 1.5 fm$^{-1}$) where they are expected to dominate and subsequently extrapolated down to low values of $q$, incurring only minor uncertainties. On the other hand, it is bad because clearly one will not be able to distinguish isospin mixing from strangeness content without another type of input. Possibly PV electron-proton scattering will be done to higher precision and the strangeness content better defined; or improved studies of elastic scattering from $^4$He can be made and used to fix the electric strangeness content without much interference from isospin mixing. Concerning the latter idea, in \cite{ram97} the size of the isospin-mixing deviation in $^4$He is estimated to be 0.3$\%$ even for a momentum transfer of $q=$ 1.5 fm$^{-1}$. The estimation of the isospin-mixing effect in that work, as well as in \cite{don89} for other $N=Z$ nuclei including $^{12}$C, is different from the one performed here. It was based on a Coulomb perturbation of the nominal $T=0$ ground state that effectively introduces a mixing with an excited $T=1$ state, both 0$^+$. The mixing parameter is proportional to the amplitude of the Coulomb perturbation and inversely proportional to the energy difference between the mixed non-perturbed states; the corresponding PV asymmetry is also dependent on the form factors of the elastic transition on the new ground state, which involves inelastic transitions in the basis previous to the perturbation. The isospin-mixing deviation for $^4$He obtained in this manner in \cite{ram97} is considered there as a conservative estimation, and the only source of isospin mixing taken into account in those works, as well as in this one, is the Coulomb interaction between protons.

Precise asymmetry measurements with this target, ideally for a set of different momentum transfers, would help to pin down the strangeness deviation uncertainty. The HAPPEX-He experiment \cite{ach07} measured the PV asymmetry with a 4$\%$ precision at $q=$ 1.4 fm$^{-1}$ (see Fig. \ref{dev_strange}). Prospects are thus good that with new techniques this level of precision can be pushed down to the desired level of 2$\%$ or better at this momentum transfer region.

\section{Conclusions}\label{conclusions}

We have analyzed in this work the sizes and uncertainties of several contributions to the PV asymmetry in polarized electron scattering by $^{12}$C related to the structure of the target. These contributions include distortion of the projectile wave functions due to the Coulomb nuclear field, isospin mixing in the nuclear states, strangeness content of the nucleons, inelastic scattering and the effect of meson-exchange currents. These are analyzed in terms of the induced deviation of the PV asymmetry with respect to the tree-level Standard Model value. The determination of other effects, such as dispersive, higher-order or box-diagrammatic corrections can be considered an important goal of the new precision experiments.
With this in mind, we have checked the statistical uncertainty expected from the nuclear effects and strangeness in the kinematic regions of interest.
We will summarize here the sizes and theoretical uncertainties these incur at the momentum transfer region of interest, centered at 0.5 fm$^{-1}$. First, Coulomb distortion effects are of a 3$\%$ size and their theoretical uncertainty has been evaluated here as 0.01 $\%$, obtained from reasonable variations of the nuclear charge distribution that provides the distorting field. Secondly, the isospin mixing in the nuclear ground state of pure electromagnetic origin accounts for a 0.4$\%$ average effect, with an uncertainty of 0.1$\%$ estimated using different nucleon-nucleon interactions in Hartree and Hartree-Fock mean field calculations. Thirdly, the current experimental knowledge of the nucleon strangeness content results in asymmetry deviations up to 1$\%$, which is also the uncertainty attached to the effect since the experimental ranges of the content parameters are nearly compatible with zero.
Other effects, such as from meson-exchange currents within the nuclear target has been addressed and estimated to be below 0.1$\%$, with the same degree of uncertainty due to the fact that this effect modifies the asymmetry only through an interplay with the isospin-mixing and strangeness contributions, provided they are present. A similar interplay may take place when excited nuclear states are reached in inelastic scattering processes, which is only an issue when their excitation energies are smaller than the experimental energy resolution. We have shown, however, that the modification to the PV asymmetry induced in $^{12}$C by the two excited states below 9.6 MeV is zero or very small.
According to the above, the strangeness content and the isospin mixing are the main sources of theoretical uncertainties concerning the analysis of the PV asymmetry in $^{12}$C. New experimental information could help improve the situation, and our suggestion in this respect involves measurements with other nuclear targets with a reduced isospin-mixing effect, as well as in other kinematic regions where one expects to have a larger strangeness content contribution. 

\begin{acknowledgments}
We acknowledge K. Kumar for fruitful discussions on experimental issues, J. M. Udias for discussions on relativistic mean field and Coulomb distortion calculations and P. Sarriguren for discussions on Skyrme interactions. This research was supported by a Marie Curie International Outgoing Fellowship within the 7th European Community Framework Programme (O. Moreno). Also supported in part (T. W. Donnelly) by the US Department of Energy under cooperative agreement DE-FC02-94ER40818. We also thank the hospitality and financial support of the MIT - LNS (Cambridge MA, US) workshop on `Polarized Electron Beams' and of the MITP-PRISMA (Johannes Gutenberg Univ. Mainz, Germany) workshop on `Low Energy Precision Physics'.
\end{acknowledgments}


\begin{thebibliography}{13}

\bibitem{MITworkshop}
MIT Workshop to Explore Physics Opportunities with Intense, Polarized Electron Beams up to 300 MeV, \emph{AIP Conf. Proc.}, to be published.

\bibitem{aul11}
K. Aulenbacher, \emph{Hyperfine Interact.} \textbf{200}, 2011, pp. 3-7.

\bibitem{nei00}
G. R. Neil \emph{et al.}, \emph{Phys. Rev. Lett.} \textbf{84}, 2000, pp- 662-665; D. Douglas \emph{et al.}, Proceedings of the 2001 Particle Accelerator Conference, Chicago, 2001, pp. 249-252.

\bibitem{don89}
T. W. Donnelly, J. Dubach and I. Sick, \emph{Nucl. Phys. A} \textbf{503}, 1989, pp. 589-631.

\bibitem{fei75}
G. Feinberg, \emph{Phys. Rev. D} \textbf{12}, 1975, pp. 3575-3582; J. D. Walecka, \emph{Nucl. Phys. A} \textbf{285}, 1977, pp. 349-367.

\bibitem{gon13}
R. Gonzalez-Jimenez, J. A. Caballero and T. W. Donnelly, \emph{Physics Reports} \textbf{524}, 2013, pp. 1-35.

\bibitem{ruf82}
G. Rufa, \emph{Nucl. Phys. A} \textbf{384}, 1982, pp. 273-286.

\bibitem{jag74}
C. W. de Jager, H. de Vries, and C. de Vries, \emph{Atomic Data and Nuclear Data Tables} \textbf{14}, 1976, pp. 479-508.

\bibitem{ang13}
I. Angeli and K. P. Marinova, \emph{Atomic Data and Nuclear Data Tables} \textbf{99}, 2013, pp. 69-95.

\bibitem{mor09}
O. Moreno, P. Sarriguren, E. Moya de Guerra, J. M. Udias, T. W. Donnelly, and I. Sick, \emph{Nucl. Phys. A} \textbf{828}, 2009,
pp. 306-332.

\bibitem{sky56}
T. H. R. Skyrme, \emph{Phil. Mag.} \textbf{1}, 1956, pp. 1043-1054; \emph{Nucl. Phys.} \textbf{9}, 1959, pp. 615-634.

\bibitem{vau}
D. Vautherin and D. M. Brink, \emph{Phys. Rev. C} \textbf{5}, 1972, pp. 626-647; D. Vautherin, \emph{Phys. Rev. C} \textbf{7}, 1973, pp. 296-316.

\bibitem{car87}
J. Carlson, \emph{Phys. Rev. C} \textbf{36}, 1987, pp. 2026-2033.

\bibitem{cha98}
E. Chabanat, P. Bonche, P. Haensel, J. Meyer, and R. Schaeffer, \emph{Nucl. Phys. A} \textbf{635}, 1998, pp. 231-256.

\bibitem{dut12}
M. Dutra, O. Lourenco, J. S. Sa Martins, A. Delfino, J. R. Stone, and P. D. Stevenson, \emph{Phys. Rev. C} \textbf{85}, 2012, 035201 pp. 1-36.

\bibitem{rein86}
P. G. Reinhard, M. Rufa, J. Maruhn, W. Greiner, and J. Friedrich, \emph{Z. Phys. A} \textbf{323}, 1986, pp. 13-26.

\bibitem{fyn05}
H. O. U. Fynbo \emph{et al.}, \emph{Science} \textbf{433}, 2005, pp. 136-139.

\bibitem{fuk13}
Y. Fukuoka, S. Shinohara, Y. Funaki, T. Nakatsukasa, and K. Yabana, \emph{Phys. Rev. C} \textbf{88}, 2013, 014321, pp. 1-14.

\bibitem{fir99}
R. B. Firestone, \emph{Table of Isotopes}, John Wiley and Sons, 1999.

\bibitem{flanz}
J. B. Flanz, \emph{PhD Thesis}, Massachusetts University Amherst (1979), unpublished.

\bibitem{dub76}
J. Dubach, J. H. Koch, and T. W. Donnelly, \emph{Nucl. Phys. A} \textbf{271}, 1976, pp. 279-316.

\bibitem{msd}
M. J. Musolf, R. Schiavilla, and T. W. Donnelly, \emph{Phys. Rev. C} \textbf{50}, 1994, pp. 2173-2188.

\bibitem{mus93}
M. J. Musolf and T. W. Donnelly, \emph{Phys. Lett. B} \textbf{318}, 1993, pp. 263-267.

\bibitem{mus94}
M. J. Musolf, T. W. Donnelly, J. Dubach, S. J. Pollock, S. Kowalski, and E. J. Beise, \emph{Physics Reports} \textbf{239}, 1994, pp. 1-178.

\bibitem{ram97}
S. Ramavataram, E. Hadjimichael, and T. W. Donnelly, \emph{Phys. Rev. C} \textbf{50}, 1994, pp. 1175-1179.

\bibitem{ach07}
A. Acha \emph{et al.}, \emph{Phys. Rev. Lett.} \textbf{98}, 2007, 032301 pp. 1-5.


\end{thebibliography}
\end{document}